\input{aipcheck}
\documentclass[final]{aipproc} 
\usepackage{graphicx,amsmath}
\layoutstyle{8x11double}
\begin{document}

\title{Supercooled liquids under shear: A mode-coupling theory approach}

\author{Kunimasa Miyazaki}{
  address={Department of Chemistry and Chemical Biology,
Harvard University, 12 Oxford Street, Cambridge, MA 02138, U.S.A}
}

\author{Ryoichi Yamamoto}{
  address={Department of Physics, Kyoto University, Kyoto 606-8502, 
           Japan},
  altaddress={PRESTO, Japan Science and Technology Agency,
              4-1-8 Honcho Kawaguchi, Saitama, Japan.} 
}

\author{David R. Reichman}{
  address={Department of Chemistry and Chemical Biology,
Harvard University, 12 Oxford Street, Cambridge, MA 02138, U.S.A}
}

\begin{abstract} 
We generalize the mode-coupling theory of supercooled fluids to 
systems under stationary shear flow. 
Our starting point is the generalized fluctuating hydrodynamic equations
with a convection term. 
The method is applied to a two dimensional colloidal suspension.
The shear rate dependence of the intermediate scattering function and 
shear viscosity is analyzed. 
The results show
a drastic reduction of the structural relaxation time due to shear and 
strong shear thinning behavior of the viscosity which 
are in qualitative agreement with recent simulations. 
The microscopic theory with minimal assumptions
can explain the behavior far beyond the linear response regime.
\end{abstract}
\maketitle

Many complex fluids such as suspensions, polymer solutions, and granular 
fluids exhibit very diverse rheological behavior. 
Shear thinning is
among the most-known phenomena. 
Recently, it was found by experiments\cite{simmons1982}
 and simulations\cite{yamamoto1998c}
that supercooled liquids near the glass-transition
also show strong shear thinning behavior. 
They have observed that near the transition temperature, the
structural relaxation time 
and the shear viscosity both decrease
as $\dot{\gamma}^{-\nu}$, where $\dot{\gamma}$ is the shear rate and
$\nu$ is an exponent which is less than but close to 1.
For such systems driven far from equilibrium, the nonequilibrium
parameter $\dot{\gamma}$ is not a small perturbation
parameter but plays a role more like an intensive parameter 
which
characterizes the ``thermodynamic state'' of the system\cite{liu1998b}.  
Such rheological behavior is
interesting in its own right, but understanding the dynamics of
supercooled liquids in a nonequilibrium state is more important because
it has possibility to shed light on an another typical and perhaps more 
important nonequilibrium problem, non-stationary aging. 
Aging is the slow relaxation after a sudden
quench of temperature below the glass transition temperature.
In this case, the waiting time plays a similar role to 
(the inverse of) the shear rate. 
Aging behavior has been extensively studied for spin 
glasses (see Ref.\cite{cugliandolo2003} and references therein).
The relationship between aging and a system driven
far away from the equilibrium was considered using 
a schematic model based on the exactly solvable p-spin spin glass 
by Berthier, Barrat and Kurchan\cite{berthier2000b}  
and its validity was tested numerically for supercooled 
liquids\cite{berthier2002}. 
There are attempts to study the aging of structural glasses 
theoretically\cite{latz2002} 
but it has not been analyzed and compared with the simulation 
results\cite{kob2000}. 

In this paper, we investigate the 
dynamics of supercooled liquids under shear
theoretically, by extending the standard mode-coupling theory (MCT).
We start with generalized fluctuating hydrodynamic equations with 
a convection term. 
Using several approximations,
we obtain a closed nonlinear equation for the 
intermediate scattering function for the sheared system.  
The theory is applicable to  both normal liquids and 
colloidal suspensions in the absence of hydrodynamic
interactions.  
Numerical results will be presented only for the 
colloidal suspensions, 
but generalization to liquids are straightforward. 
Some of the preliminary results have already been published in 
Ref.\cite{miyazaki2002}. 

We shall consider a two dimensional colloidal suspension under 
a stationary simple shear flow given by 
\begin{equation}
{\bf v}_{0}({\bf r}) = \mbox{\boldmath$\Gamma$}\cdot{\bf r} 
= (\dot{\gamma} y, 0), 
\end{equation}
where 
$(\mbox{\boldmath$\Gamma$})_{\alpha\beta}
=\dot{\gamma}\delta_{\alpha x}\delta_{\beta y}$
is the velocity gradient matrix.
The hydrodynamic fluctuations for density $\rho({\bf r}, t)$ and the
velocity field ${\bf v}({\bf r}, t)$ obey the following set of Langevin
equations\cite{kirkpatrick1986c}.  
\begin{equation}
\begin{split}
&
\frac{\partial{\rho}}{\partial t}
= - \nabla\cdot(\rho{\bf v}),
\\
&
m\frac{\partial (\rho{\bf v})}{\partial t}
+m\nabla\cdot(\rho{\bf v}{\bf v}) 
= 
-\rho \nabla\frac{\delta {\cal F}}{\delta \rho}
-\zeta_{0}\rho({\bf v}-{\bf v}_{0}) + {\bf f}_{R},
\end{split}
\label{eq:hydrodynamcs}
\end{equation}
where $\zeta_{0}$ is the collective friction coefficient for colloidal
particles. 
${\bf f}_{R}({\bf r}, t)$ is the random force which satisfies
the fluctuation-dissipation theorem of the second kind (2nd FDT);
\begin{equation}
\langle
{\bf f}_{R}({\bf r}, t){\bf f}_{R}({\bf r'}, t')
\rangle_{0}
= 2k_{\mbox{\tiny B}}T\rho({\bf r})\zeta_{0}
\delta({\bf r}-{\bf r}')\delta(t-t')
\end{equation}
for $t\geq t'$, 
where $\langle\cdots\rangle_{0}$ is an average over the conditional
probability for a fixed value of $\rho({\bf r})$ at $t=t'$. 
Note that the random force depends on the density and thus
the noise is multiplicative. 
We assumed that the 2nd FDT holds even in nonequilibrium
state since the correlation of the random forces are short-ranged and
short-lived, and thus the effect of the shear is expected to 
be negligible. 
The friction  term is specific for the colloidal case. 
In the case of liquids, it should be replaced by 
a stress term which is proportional to the gradient of the 
velocity field
multiplied by the shear viscosity.
Both cases, however, lead to the same dynamical behavior on long time
scales.
The first term in the right hand side of the equation for the
momentum is the pressure term and ${\cal F}$ is the total free energy
in a stationary state.  Here we assume that the free energy is well
approximated by that of the equilibrium form and is given by a
well-known expression;
\begin{equation}
\begin{split}
\beta{\cal F} 
\simeq 
&
\int\!\!\mbox{d}{\bf r}~ \rho({\bf r})\left\{ \ln\rho({\bf r})/\rho_0 -1 \right\}
\\
&
-\frac{1}{2}\int\!\!\mbox{d}{\bf r}_1\int\!\!\mbox{d}{\bf r}_2~
\delta\rho({\bf r}_1)c(r_{12})\delta\rho({\bf r}_2),
\end{split}
\end{equation}
where $\beta=1/k_{\mbox{\tiny B}} T$ and 
$c(r)$ is the direct correlation function.
Under shear, it is expected that $c(r)$ will be distorted and should
be replaced by a nonequilibrium, steady state form
$c_{\mbox{\tiny NE}}({\bf r})$, which is an anisotropic function of
${\bf r}$.  
It is, however, natural to expect that this distortion is very small in
the molecular length scale, which plays the most important role in the
slowing down of the structural relaxation near the glass transition. 
We  confirmed this by numerical simulation\cite{miyazaki2003prep}.
By linearizing eq.(\ref{eq:hydrodynamcs}) around the stationary state as
$\rho = \rho_{0}+\delta \rho$ and ${\bf v} = {\bf v}_{0} + \delta{\bf v}$, where
$\rho_{0}$ is the average density, we obtain the following equations,
\begin{equation}
\begin{split}
&
\left(
\frac
{\partial~}{\partial t}- {\bf k}\cdot\mbox{\boldmath$\Gamma$}\cdot
\frac{\partial ~}{\partial {\bf k}}
\right)
\delta\rho_{{\bf k}}(t) = -ik J_{{\bf k}}(t),
\\
&
\left(
\frac{\partial ~}{\partial t}
- {\bf k}\cdot\mbox{\boldmath$\Gamma$}\cdot
 \frac{\partial ~}{\partial {\bf k}}
+ \hat{\bf k}\cdot\mbox{\boldmath$\Gamma$}\cdot\hat{\bf k}
\right)
J_{{\bf k}}(t)
\\
&
= 
-\frac{ik}{m\beta S(k)}\delta\rho_{{\bf k}}(t)
\\
&
-\frac{1}{m\beta}
\int_{{\bf q}}i\hat{\bf k}\cdot{\bf q} c(q)\delta\rho_{{\bf k}-{\bf q}}(t)\delta\rho_{{\bf q}}(t)
-\frac{\zeta_{0}}{m}J_{{\bf k}}(t) + {\bf f}_{R{\bf k}}(t),
\end{split}
\label{eq:nonlinear}
\end{equation}
where 
$c(q)$ is the Fourier transform
of $c(r)$, 
$\hat{\bf k} \equiv {\bf k}/|{\bf k}|$,
$J_{{\bf k}}(t)=\rho_{0}\hat{\bf k}\cdot\delta{\bf v}_{{\bf k}}(t)$ is the
longitudinal momentum fluctuation, 
and 
$\int_{{\bf q}}\equiv \int\mbox{d}{\bf q}/(2\pi)^2$.
Note that our approximate equation does not contain
coupling to transverse momentum fluctuations even in the presence of
shear.

In order to construct equations for the appropriate correlations
from the above expressions, an approximate symmetry is necessary.  In
the presence of shear, translational invariance is violated.  In other
words, correlations of arbitrary fluctuations, $f({\bf r},t)$ and $g({\bf r},
t)$, do not satisfy $\langle f({\bf r},t)g({\bf r}', 0)\rangle \neq \langle
f({\bf r}-{\bf r}',t)g({\bf 0},0)\rangle$.  
Instead, it has the following symmetry\cite{onuki1997};
\begin{equation}
\langle f({\bf r}, t)g({\bf r}', 0) \rangle
= 
\langle f({\bf r}-{\bf r}'(t), t)g({\bf 0}, 0) \rangle,
\label{eq:transinv}
\end{equation}
or in wavevector space
\begin{equation}
\begin{split}
\langle f_{\bf k}(t)g^{*}_{{\bf k}'}(0) \rangle
&
= 
\langle f_{{\bf k}}(t)g^{*}_{{\bf k}(t)}(0) \rangle
\times \delta_{{\bf k}(t),{\bf k}'},
\end{split}
\label{eq:inv}
\end{equation}
where we defined the time-dependent position and wave vector by 
${\bf r}(t) \equiv \exp[\mbox{\boldmath$\Gamma$} t]\cdot{\bf r} 
= {\bf r} + \dot{\gamma} t y \hat{\bf e}_{x}$, where 
$\hat{\bf e}_{x}$ is an unit vector oriented along the $x$-axis
and 
${\bf k}(t) 
= \exp[{}^{t}\mbox{\boldmath$\Gamma$} t]\cdot{\bf k} = {\bf k} + \dot{\gamma} t
k_{x}\hat{\bf e}_{y}$, 
where ${}^{t}\mbox{\boldmath$\Gamma$}$ denotes the transpose of
$\mbox{\boldmath$\Gamma$}$ and $\delta_{{\bf k},{\bf k}'}\equiv
(2\pi)^{2}V^{-1}\delta({\bf k}-{\bf k}')$ for a system of 
volume $V$.
Using this approximation, it is straightforward to construct the
mode-coupling equations for the appropriate correlation functions. 
We shall derive the equation for the intermediate scattering function 
defined by 
$F({\bf k}, t) \equiv 
{N}^{-1}\langle \delta\rho_{{\bf k}(-t)}(t)\delta\rho^{*}_{{\bf k}}(0) \rangle$,
where $N$ is the total number of the particles in the system.
Note that 
the wave vector in $\delta\rho_{{\bf k}}(t)$ is now replaced  
by a time-dependent one ${\bf k}(-t)$.

Eq.(\ref{eq:nonlinear}) has a quadratic nonlinear term in 
$\delta\rho_{{\bf k}}(t)$. 
This term can be renormalized 
to give a generalized friction coefficient or the memory kernel
following the standard procedure of derivation of the 
mode-coupling equation\cite{martin1973}. 
To the lowest order in the loop expansions, we obtains the equation for 
the velocity-density correlation
$C({\bf k}, t) \equiv N^{-1}
\langle J_{{\bf k}(-t)}(t) n^{*}_{{\bf k}}(0) \rangle$;
\begin{equation}
\begin{split}
&
\frac{\mbox{d} C({\bf k}, t) }{\mbox{d} t}
-\hat{\bf k}(-t)\cdot\mbox{\boldmath$\Gamma$}\cdot\hat{\bf k}(-t)C({\bf k}, t)
= 
- \frac{ik(-t)F({\bf k}, t)}{m\beta S(k(-t))}
\\
&
-\frac{1}{m}\int_{0}^{t}\!\!\mbox{d} t'~
 \int\!\!\mbox{d}{{\bf k}'}~
 \zeta({\bf k}(-t), {\bf k}', t-t')C({\bf k}', t').
\end{split}
\label{eq:C}
\end{equation}
Note that in the above equation, the differential operator 
${\bf k}\cdot\mbox{\boldmath$\Gamma$}\cdot\partial/\partial{\bf k}$ 
disappears and ${\bf k}$ is replaced by ${\bf k}(t)$.
$\zeta({\bf k}, {\bf k}', t)$ is the generalized friction coefficient.
$\zeta({\bf k}, {\bf k}', t)$ is given by the sum of the bare friction
coefficient and the mode-coupling term as
\begin{equation}
\zeta({\bf k}, {\bf k}', t)
= \zeta_{0}\times 2\delta(t)+\delta\zeta({\bf k}, t)\delta_{{\bf k}(t),{\bf k}'}
\end{equation}
with the mode-coupling contribution given by 
\begin{equation}
\begin{aligned}
\delta\zeta({\bf k}, t)
= 
&
\frac{\rho_{0}}{2\beta}
\int_{{\bf q}}
{\cal V}({\bf k}, {\bf q}){\cal V}({\bf k}(t), {\bf q}(t))
\\
&
\times
F({\bf k}(t)-{\bf q}(t), t)F({\bf q}(t), t), 
\end{aligned}
\label{eq:MCT}
\end{equation}
where ${\cal V}({\bf k}, {\bf q})$ is the vertex function given by 
\begin{equation}
{\cal V}({\bf k}, {\bf q})
=
\hat{\bf k}\cdot
\left\{
{\bf q} c({\bf q}) + \left({\bf k}-{\bf q} \right)c({\bf k}-{\bf q}) 
\right\}.
\end{equation}
Note that, in the derivation of eq.(\ref{eq:MCT}), we have assumed the 
fluctuation-dissipation theorem of the first kind (1st FDT), which
relates the response function to the correlation function.  
In order to derive the renormalized friction coefficient, 
propagators (the response function to the random forces) as well as 
correlation functions\cite{martin1973} are naturally introduced. 
The 1st FDT makes it possible to eliminate 
the propagators in favour of the correlation functions. 
In the overdamped limit, we may neglect the time-derivative of 
$C({\bf k}, t)$. 
The second term on the left hand side is also neglected if
P\'{e}clet number Pe$=\dot{\gamma}
\sigma^2/D_0$ ($\sigma$ is the diameter of the particle and
$D_0=k_{\mbox{\tiny B}} T/\zeta_0$ is the diffusion coefficient) 
is small.
Therefore, combining the first equation of eq.(\ref{eq:nonlinear}),
one may eliminate $C({\bf k}, t)$ from the above 
equations and arrive at the closed equation for $F({\bf k}, t)$; 
\begin{equation}
\begin{split}
\frac{\mbox{d} F({\bf k}, t) }{\mbox{d}t}
= 
&
-\frac{D_{0}k(-t)^2}{S(k(-t))}F({\bf k}, t)
\\
&
-\int_{0}^{t}\!\!\mbox{d} t'~
  M({\bf k}(-t),t-t')
  \frac{\mbox{d} F({\bf k}, t') }{\mbox{d}t'},
\end{split}
\label{smolchowski}
\end{equation}
where 
\begin{equation}
\begin{split}
M({\bf k},t)
= 
&
\frac{\rho_{0}D_{0}}{2}
\frac{k}{k(t)}
\int_{{\bf q}}
{\cal V}({\bf k}, {\bf q}){\cal V}({\bf k}(t), {\bf q}(t))
\\
&
\times
F({\bf k}(t)-{\bf q}(t), t)F({\bf q}(t), t).
\end{split}
\label{eq:Mkt}
\end{equation}
In the absence of the shear, they reduce to the conventional 
mode-coupling equation\cite{gotze1992}.  

Solving 
eqs.(\ref{smolchowski}) and (\ref{eq:Mkt}) numerically
is more demanding than the 
equations in the equilibrium state because the wavevectors are
distorted by shear and the system is not isotropic.  
We have considered a two dimensional colloidal suspension consisting of
hard disks. 
For the static correlation function $c(k)$ and $S(k)$, 
the analytic expressions derived by Baus {\sl et al.}\cite{baus1987}
were used. 
We have divided the two dimensional wavevector space into $N_k$ grids
for each direction. 
The cut-off wavevector was chosen to be $k_c\sigma = 10\pi$.
For the self-consistent calculation of the mode-coupling equation, 
we have used the algorithm developed by Fuchs {\sl et
al.}\cite{fuchs1991}. 
In the following results, we have used the grid number $N_k=55$, for which
the ergodic-nonergodic transition occurs at the volume fraction
$\phi_c = \pi \sigma^2 \rho_{0}/4 = 0.7665$. 
Apparently, $N_k$ is not large enough to give the right 
transition density which is 10\% smaller. 
However, the qualitative behaviors does not change by increasing 
the grid number.

In Figure 1, we show the behavior of $F({\bf k}, t)$ 
for $\phi= \phi_c \times (1 - 10^{-4})$ for various 
shear rate Pe=$10^{-10}$ to $10$.   
\begin{figure}[ht]
\includegraphics[scale=0.62,angle=0]{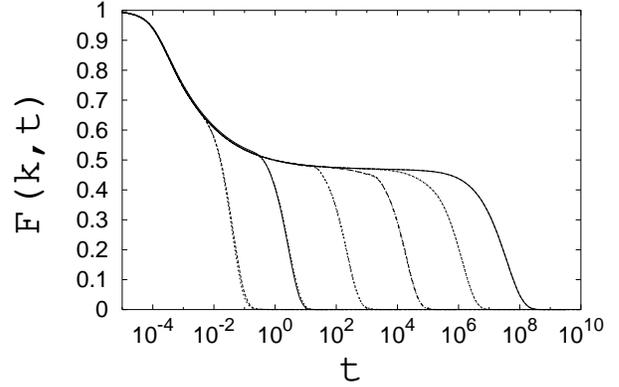} 
\caption{$F({\bf k}, t)/S(k)$ for ${\bf k}\sigma=(0, 3),~(3, 0)$ 
for various 
shear rates $\dot{\gamma}$.
From the right to the left, 
$\mbox{Pe} =10^{-10}$, $10^{-7}$, 
$10^{-5}$, $10^{-3}$, $10^{-1}$, and $10$.
The time $t$ is scaled by $\sigma^2/D_{0}$. 
} 
\label{fig:1}
\end{figure}
The wavevectors were chosen to be ${\bf k}\sigma = (0, 3)$ and $(3,0)$,
parallel and perpendicular to the shear flow, respectively.
For the smaller shear rate, $\mbox{Pe} < 10^{-10}$, 
we did not see any shear effect. 
For higher shear rates, 
we have observed the drastic reduction of the relaxation time
due to shear. 
This is similar to the behavior reported
in recent molecular dynamics 
simulations for a binary soft-core liquid\cite{yamamoto1998c}.  

The two lines for a fixed shear magnitude but for wavevectors
along different directions
collapsed onto each other 
and we do not see
a noticeable difference. 
It is surprising that, though the perturbation of shear flow
is highly
anisotropic, the dynamics of fluctuations are almost isotropic. 
This fact is also observed in the simulation for a binary
liquid\cite{miyazaki2003prep}. 
The reason for the isotropic nature is 
understood as follows. 
The shear flow perturbs and
randomizes the phase of coupling between different modes.  
This perturbation dissipates the cage that transiently immobilizes
particles.  
Mathematically, this is reflected through the time
dependence of the vertex.
This ``phase randomization'' occurs irrespective of the direction of the
wavevector, which results in the isotropic behavior of relaxation. 
This mechanism is very different from 
that of many complex fluids and dynamic critical
phenomena under shear, in which the faster relaxation occurs mainly 
due to the distortion of the structures at small wavevectors which are
stretched out by the shear flow and
pushed to larger wavevectors where faster relaxation occurs.

The shear viscosity, $\eta$, is easily evaluated by modifying the 
Green-Kubo formula for the sheared system\cite{kirkpatrick1985};
\begin{equation}
\begin{aligned}
&
\eta(\dot{\gamma})
=
\eta_0
\\
&
+
\frac{1}{2\beta}
\int_{0}^{\infty}\!\!\mbox{d} t
\int_{{\bf k}}
\frac{k_{x}k_{x}(t)}{S^2(k)S^2(k(t))}
\frac{\partial S(k)}{\partial k_{y}}
\frac{\partial S(k(t))}{\partial k_{y}(t)}
F^2({\bf k}(t),t)
,
\end{aligned}
\end{equation}
where $\eta_{0}$ is the viscosity of the solvent alone. 
The integral can be implemented for the set of $F(k,t)$ evaluated using 
eq.(\ref{smolchowski}). 
Shear rate dependence of the reduced viscosity 
$\eta_{R}(\dot{\gamma})\equiv \{ \eta(\dot{\gamma}) -\eta_{0}\}/\eta_{0}$ 
is plotted
in Figure 2 for various densities around $\phi_c$.
\begin{figure}[ht]
\includegraphics[scale=0.62,angle=0]{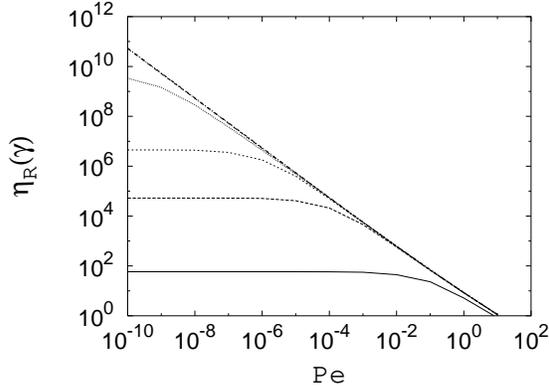} 
\caption{The reduced viscosity 
is plotted for $\mbox{Pe} =\dot{\gamma}\sigma^2/D_{0}$ for various densities.  
From above to bottom;
$\phi = 0.766549$, $0.7665$, $0.766453$, $0.7664$, $0.766$, and $0.756$.
The highest density is $4\times 10^{-5}$ \% larger than $\phi_c$. 
} 
\label{fig:3}
\end{figure}
The strong non-Newtonian behavior is observed at high shear rate and
large densities, which is again in qualitative agreement with the
simulation results for liquids. 
Slightly above $\phi_c$, the plastic behavior which implies the presence
of the yield stress is also observed.
The shear thinning exponent is extracted from this plot between 
$10^{-10}~ < ~\mbox{Pe} ~ < 1$
and we obtained
$\eta_{R}(\dot{\gamma})\propto \dot{\gamma}^{-\nu}$ with $\nu \simeq 
0.99$. 
For the larger shear rate, Pe $ > 1$, the exponent becomes smaller. 
In this regime, it is expected that other mechanism 
such as the distortion of structure $c(k)$ and $S(k)$ by shear becomes
important. 

The mode-coupling theory developed in this 
paper is far from complete. 
The most crucial approximation is the use of the 1st FDT, which was
employed when we close the equation in terms of the correlation
functions alone. 
It is already known that the 1st FDT is violated for supercooled 
systems
under shear as well as during aging\cite{berthier2002}. 
Without the 1st FDT, one has to solve simultaneously
the set of mode-coupling equations
for the propagator and correlation function, which couple each other
through the memory kernels. 
Research in this direction is under way.
Another important approximation was to neglect 
the small distortion of the structure, $c(k)$ and $S(k)$, due to
shear. 
The construction of the equation for such an equal-time correlation
functions might be more subtle and should be considered in future.
It is surprising, however, that despite of these approximations, 
the theory reproduces the major 
features which was seen in simulations;
the drastic 
reduction of the relaxation time, the isotropic nature of the dynamics,
and plastic-like strong shear thinning. 

An analogous effort has been made by Fuchs and Cates\cite{fuchs2002c}. 
They have derived a mode-coupling expression for $F(k,t)$
using a projection operator for the Smoluchowski equation for
$N$-particle colloidal suspensions. 
They have observed similar shear thinning behavior for
approximated expressions, for an isotropic model, where 
the anisotropy hidden in the equations are neglected. 

The details of analytical and numerical calculations for results given
in the present paper are given elsewhere\cite{miyazaki2003prep}.

\begin{theacknowledgments}
The authors acknowledge support from NSF grant \#0134969.  The authors
would like to express their gratitude to Prof. Matthias Fuchs for
for useful discussions and helpful instruction about the program codes 
to solve the mode-coupling equation.
\end{theacknowledgments}

\bibliographystyle{aipproc}

\begin{thebibliography}{20}
\expandafter\ifx\csname natexlab\endcsname\relax\def\natexlab#1{#1}\fi
\providecommand{\enquote}[1]{``#1''}
\expandafter\ifx\csname url\endcsname\relax
  \def\url#1{\texttt{#1}}\fi
\expandafter\ifx\csname urlprefix\endcsname\relax\def\urlprefix{URL }\fi

\bibitem[Simmons et~al.(1982)]{simmons1982}
Simmons, J.~H., Mohr, R.~K., and Montrose, C.~J., \emph{J. Appl. Phys.},
  \textbf{{\bf 53}}, 4075 (1982).

\bibitem[Yamamoto and Onuki(1998)]{yamamoto1998c}
Yamamoto, R., and Onuki, A., \emph{Phys. Rev. {\rm E}}, \textbf{{\bf 58}},
  3515 (1998).

\bibitem[Liu and Nagel (1998)]{liu1998b}
Liu, A.~J., and Nagel, S.~R., \emph{Nature}, \textbf{{\bf 396}}, 21
  (1998).

\bibitem[Cugliandolo (2003)]{cugliandolo2003}
Cugliandolo, L.~F., in \emph{{"Slow
  relaxations and nonequilibrium dynamics in condensed matter"}}, ed.
  J.-L. Barrat, M.~Feigelman, and J.~Kurchan, Springer-Verlag, New York,
  (2003), pp. 371.

\bibitem[Berthier et~al. (2000)]{berthier2000b}
Berthier, L., Barrat, J.-L., and Kurchan, J., \emph{Phys. Rev. {\rm E}},
  \textbf{{\bf 61}}, 5464 (2000).

\bibitem[Berthier and Barrat (2002)]{berthier2002}
Berthier, L., and Barrat, J.-L., \emph{Phys. Rev. Lett.}, \textbf{{\bf
  89}}, 095702 (2002).

\bibitem[Latz (2002)]{latz2002}
Latz, A., \emph{J. Stat. Phys.}, \textbf{{\bf 109}}, 607
  (2002).

\bibitem[Kob et~al. (2000)]{kob2000}
Kob, W., Barrat, J.-L., Sciortino, F., and Tartaglia, P., \emph{J. of Phys.:
  Condens. Matter}, \textbf{{\bf 12}}, 6385 (2000).

\bibitem[Miyazaki and Reichman (2002)]{miyazaki2002}
Miyazaki, K., and Reichman, D.~R., \emph{Phys. Rev. {\rm E}}, \textbf{{\bf
  66}}, 050501(R) (2002).

\bibitem[Kirkpatrick and Nieuwoudt(1986)]{kirkpatrick1986c}
Kirkpatrick, T.~R., and Nieuwoudt, J.~C., \emph{Phys. Rev. {\rm A}},
  \textbf{{\bf 33}}, 2651 (1986).

\bibitem[Miyazaki et~al.(unpublished)]{miyazaki2003prep}
Miyazaki, K., Yamamoto, R., and Reichman, D.~R., (unpublished).

\bibitem[Onuki(1997)]{onuki1997}
Onuki, A., \emph{J. of Phys.: Condens. Matter}, \textbf{{\bf 9}}, 6119
  (1997).

\bibitem[Martin et~al.(1973)]{martin1973}
Martin, P.~C., Siggia, E.~D., and Rose, H.~A., \emph{Phys. Rev. {\rm A}},
  \textbf{{\bf 8}}, 423 (1973).

\bibitem[G{\"{o}}tze and Sj{\"{o}}gren (1992)]{gotze1992}
G{\"{o}}tze, W., and Sj{\"{o}}gren, L., \emph{Rep. Prog. Phys.},
  \textbf{{\bf 55}}, 241 (1992).

\bibitem[Baus and Colot(1987)]{baus1987}
Baus, M., and Colot, J.~L., \emph{Phys. Rev. {\rm A}}, \textbf{{\bf 36}},
  3912 (1987).

\bibitem[Fuchs et~al.(1991)]{fuchs1991}
Fuchs, M., {G\"{o}tze}, W., and Latz, A., \emph{J. of Phys: Condens.
  Matter}, \textbf{{\bf 3}}, 5047 (1991).

\bibitem[Kirkpatrick(1985)]{kirkpatrick1985}
Kirkpatrick, T.~R., \emph{J. of Non-Crystalline Solids}, \textbf{{\bf 75}},
  437 (1985).

\bibitem[Fuchs and Cates (2002)]{fuchs2002c}
Fuchs, M., and Cates, M.~E., \emph{Phys. Rev. Lett.}, \textbf{{\bf 89}},
  248304 (2002).

\end{thebibliography}

\end{document}